\documentclass[aps,pra,superscriptaddress,epsfigure,twocolumn]{revtex4-1}
\usepackage{graphicx}
\usepackage{amsmath}    
\usepackage{latexsym}
\usepackage{amsfonts}   
\usepackage{amssymb}
\usepackage{array}      
\usepackage{epsfig}
\usepackage{txfonts}
\usepackage{hyperref}
\usepackage{color}
\usepackage{braket}
\usepackage{float}
\usepackage{mathrsfs}
\usepackage{mathtools}

\DeclareMathOperator{\tr}{Tr}

\begin{document}

\title{Quantum correlations and thermodynamic performances of two-qubit engines \\ with local and common baths}

\author{Adam Hewgill}
\affiliation{Centre  for  Theoretical  Atomic,  Molecular  and  Optical  Physics, Queen's  University  Belfast,  Belfast  BT7 1NN,  United  Kingdom}
\author{Alessandro Ferraro}
\affiliation{Centre  for  Theoretical  Atomic,  Molecular  and  Optical  Physics, Queen's  University  Belfast,  Belfast  BT7 1NN,  United  Kingdom}\author{Gabriele De Chiara}
\affiliation{Centre  for  Theoretical  Atomic,  Molecular  and  Optical  Physics, Queen's  University  Belfast,  Belfast  BT7 1NN,  United  Kingdom}
\affiliation{Kavli Institute of Theoretical Physics (KITP), University of California, Santa Barbara CA 93106-4030, United States of America}

\begin{abstract}
 We investigate heat engines whose working substance is made of two coupled qubits performing a generalised Otto cycle by varying their applied magnetic field or their interaction strength during the compression and expansion strokes. During the heating and cooling strokes, the two qubits are coupled to local and common environments that are not necessarily at equilibrium. We find  instances of quantum engines coupled to non equilibrium common environments exhibiting non-trivial connections to quantum correlations as witnessed by a monotonic dependence of the work produced on quantum discord and entanglement.
\end{abstract}

\maketitle
\section{Introduction}
Quantum thermodynamics is an active area of research that focuses on concepts derived from classical thermodynamics---like heat, work,  and the laws of thermodynamics---and aims at  understanding and exploiting them in the quantum context \cite{0305-4470-12-5-007, e15062100,LIEB19991,GooldReview,AndersReview,XuerebReview}. Traditionally, considerable efforts have focused on introducing quantum versions of classical engines \cite{PhysRevLett.2.262}. A wide range of quantum heat engines \cite{PhysRevE.76.031105} have been devised, notably based on the Otto \cite{PhysRevE.83.031135,1402-4896-88-6-065008,Zhao2017,KarimiPRB2017,PhysRevE.86.051105,PhysRevE.90.032102, PhysRevE.94.012137,  PhysRevE.94.012137, e19040136, PhysRevE.92.022142, PhysRevE.79.041113, PhysRevA.96.052119,PhysRevE.90.032102,Reid} and Carnot cycles \cite{Scully862,0295-5075-88-5-50003,0295-5075-117-5-50002, PhysRevE.92.012118,GardasPRE2015, 0143-0807-23-5-306,PhysRevE.89.032115}, among others \cite{Huang2014,CampisiNJP2015,PhysRevE.94.032120,NiedenzuPRE2015,Scopa2018}. In this context, a large selection of quantum working substances has been used to devise these engines, including  qubits~\cite{PhysRevE.83.031135,1402-4896-88-6-065008,Zhao2017,KarimiPRB2017,PhysRevE.86.051105,PhysRevE.90.032102,CampisiNJP2015,PhysRevE.89.032115,Huang2014}, qudits \cite{PhysRevE.79.041113,PhysRevE.92.022142,NiedenzuPRE2015,PhysRevE.94.032120,PhysRevA.96.052119,PhysRevE.92.012118,GardasPRE2015}, photons \cite{0295-5075-88-5-50003,0295-5075-88-5-50003,0295-5075-117-5-50002,Scully862} and harmonic oscillators \cite{e19040136,0143-0807-23-5-306,PhysRevE.94.012137,Reid,Scopa2018}. 
A few works  have also dealt with many-body powered quantum engines \cite{CampisiNatComm2016,JaramilloNJP2016}. Experimental platforms have also reached a level progress that allows for these heat engines to be constructed, and theoretical predictions tested and verified   \cite{2017arXiv171008716K, Rossnagel325,peterson2018experimental}. 

The proper understanding of quantum heat engines requires that the role played by the genuine quantum properties of their working substance be investigated. Among these properties, 
quantum correlations in general \cite{modi2012classical}, and entanglement in particular \cite{horodecki2009quantum}, are recognised as crucial in setting the departure of the quantum from the classical description of physical systems.
While mostly investigated  as a resource for information processing, quantum correlations have also been significantly considered in the context of many-body systems \cite{amico2008entanglement,DeChiaraROPP2018} and, more recently, in quantum thermodynamics \cite{GooldReview}.

This work aims at considering in details the relation, if any, between the quantum correlations established in the working substance during the operation of heat engines and their respective energetic performances, specifically in terms of work output and efficiency. For this, we study working substances composed of more than one constituent only, focusing in particular on the two-qubit case. While this has been partially touched on in previous works \cite{1402-4896-88-6-065008,PhysRevA.96.052119,PhysRevE.90.032102,0295-5075-88-5-50003,PhysRevE.89.032115,PhysRevLett.111.230402}, the latter have been mostly confined to reservoirs with equilibrium thermal steady states, which in turn limit the correlations that can be developed in the working substance. 

Here, we instead focus on designing engines that \textit{(i)} operate out of thermal equilibrium and \textit{(ii)} interact with baths acting either locally (namely, with each qubit separately) or commonly (namely, via simultaneous interactions with both qubits). Interestingly, this approach turns out to be rich enough to cover a variety of instances: it is possible to design case studies for which the energetic performances are either completely unrelated to quantum correlations, or related to quantum discord, or to entanglement, depending on the specific design of the engine. A  summary of the various instances we introduce is given in Tab.~\ref{tab}. 

\begin{table}[b]
\begin{tabular}{|p{1.3cm}|p{1.3cm}|p{1.3cm}|p{1.3cm}|p{1.9cm}|}
 \hline 
  & Bath & steady state & projected state & relation with work produced\\ 
 \hline 
 Sec. \ref{Loc_A} & local & separable & separable & none \\ 
 \hline 
 Sec. \ref{Loc_B} & local & entangled & separable & none \\ 
 \hline 
  Sec. \ref{Local Hamiltonian} & common & entangled & discordant & discord \\ 
 \hline 
 Sec. \ref{Common_interacting} & common & entangled & entangled & entanglement \\ 
 \hline 
 \end{tabular}  
  \caption{Summary of the cases analysed in this paper and the corresponding sections. We list whether the baths are local on the two qubits or common; whether the steady state or the state after energy projection is separable or contains quantum correlations; and finally the relation between quantum correlations and the work produced.}
   \label{tab}
   \end{table}

We employ techniques developed for open quantum systems \cite{PhysRevA.65.010101, PhysRevLett.89.277901,0295-5075-101-6-60005,JOUR, 0143-0807-33-4-805,lindblad1976,GSK76,Volovich,Gonzalez2017,HoferNJP2017} and use steady states thereof that allow us to examine a wide selection of cases, including non-equilibrium and common reservoirs. The setups  considered here can be realised experimentally by letting the working substance undergo repeated interactions with external ancillary quantum systems \cite{PhysRevE.90.042142,PhysRevLett.102.207207,BaraReaptedinteaction,PhysRevX.7.021003}   and are immune to thermodynamic inconsistencies~\cite{LevyKosloffEPL2014,ourLocalmasterequation}. For non-equilibrium reservoirs, it is important to stress that maintaining them requires extra work that we do not analyse here.

We examine two-qubit working substances with an XY Hamiltonian that undergo a generalised Otto cycle. We look into the possibility of generating quantum correlations at the steady state when using local and common jump operators. The effect of measurements necessary for a consistent definition of thermodynamic work in a quantum setting will be taken into account as well. We will also investigate the discrepancy in steady state that occurs between a local thermal master equation and a typical thermal bath, highlighting the effects on the performances of heat engines.

The paper is organised as follows. In Sec.~\ref{Model}, we introduce the working substance and the method for constructing and calculating the work and efficiency of a heat cycle. In Sec.~\ref{Engine with local baths}, we analyse two-qubit Otto engines coupled to local reservoirs, while in Sec.~\ref{Common Baths and Engine Performance.}, we expand this analysis to include a selection of common baths resulting in heat engines whose work produced depends on the quantum correlations present in the system. Finally, in Sec.~\ref{sec:summary} we summarise and discuss our findings.

\section{Model}\label{Model}
We consider as working substance a system comprised of two interacting qubits $a$ and $b$ described by the XY Hamiltonian subject to a magnetic field:
\begin{equation}\label{hamiltonian}
H= (J_x\sigma_{xa}\sigma_{xb}+J_y\sigma_{ya}\sigma_{yb})+B (\sigma_{za}+\sigma_{zb})\;,
\end{equation} 
where $\sigma_{[x,y,z]\,j}$ represents the Pauli operator acting on qubit $j=a,b$, the positive coefficients $J_x$ and $J_y$ are the strengths of the antiferromagnetic couplings and $B$ is the applied magnetic field.

The working substance can be put in contact with one or more reservoirs, not necessarily at equilibrium. This means that after the system is put in contact with the reservoir for a long time it reach not a thermal state but rather a non-equilibrium steady state.
In the following we  consider separately the case of local reservoirs, each interacting with one of the qubits, and common reservoirs interacting globally with the two-qubit system. In any case, we assume that the dynamics of the working substance density operator $\rho$ is Markovian and can be described by a Lindblad master equation of the form
\begin{equation}\label{eq:Masterequation}
\dot{\rho}= -i[H,\rho] + \sum_i g_i \mathcal{L}_{a_i}(\rho),
\end{equation}
where $[\cdot,\cdot]$ denotes the commutator, $\mathcal{L}_{a_i}=2 a_i \rho a_i^\dagger -\left\{a_i^\dagger a_i,\rho\right\}$, $a_i$ is a jump operator describing the action of the bath and $\left\{\cdot,\cdot \right\}$ denotes the anticommutator. The coefficient $g_i$ denotes the rate of dissipation associated with the Lindblad term $\mathcal{L}_{a_i}$. In the absence of interaction with the environment we  assume the evolution of the system to be described by Eq.~\eqref{eq:Masterequation} with all $g_i=0$. 

The working substance undergoes a generalised Otto cycle \cite{PhysRevE.76.031105}. The four stages of the cycle are as follows:
\begin{enumerate}
	\item \textit{Compression}: The working substance is isolated from any environment and one of the Hamiltonian's parameters, $J_x,\, J_y$ or $B$, is changed inducing an increase in the energy gaps.
	\item \textit{Heating}: The working substance is put into contact with one or more reservoirs which may be at different temperatures. During the evolution, the working substance converges to a steady state, i.e. satisfying Eq.~\eqref{eq:Masterequation} with $\dot\rho=0$. At the end of this evolution the average energy of the working substance is increased meaning that heat is absorbed from the reservoirs.
	\item \textit{Expansion}: The working substance is isolated from any environment and one of the Hamiltonian's parameters, $J_x,\, J_y$ or $B$, is changed back to its original value inducing a decrease in the energy gaps.
	\item \textit{Cooling}: The working substance is put into contact with one or more reservoirs which may be at different temperatures and allowed to reach a steady state by releasing heat into the environment. The final state is also the initial state of the  working substance at the beginning of a new cycle.
\end{enumerate}
A useful property of the Otto cycle is that during the  expansion and compression strokes, the system is isolated from the environment and thus its energy change is due only to external work. On the other hand for the heating and cooling strokes and for non equilibrium reservoirs, there might be an exchange of both heat and work. We comment on this delicate issue in every analysis we make.
 
For our Hamiltonian \eqref{hamiltonian} the parameter $P(t)$ to be changed is either $J_x,\, J_y$ or $B$ depending on what terms are present in the given Hamiltonian. We assume that such a parameter changes from its initial value $P_1$ to its final value $P_2$ via a linear ramp
\begin{equation}
P(t)=P_1+\dfrac{P_2-P_1}{\tau_{\rm ramp}}t\;,
\end{equation}
where $\tau_{\rm ramp}$ is the duration of the work stroke. 

To assess the work extracted/produced by the engine during the cycle we employ the definition of work based on the two-time measurement protocol \cite{TalknerPRE2007, CampisiRMP2011,PhysRevE.94.012137, PhysRevLett.118.050601} which we now briefly report. Let us assume the system to be initially in state $\rho^{\rm in}$ and subject to the initial Hamiltonian $H_{\rm in}$ with eigenvalues $E^{\rm in}_i$ and orthonormal eigenvectors $\ket {e^{\rm in}_{i\alpha}}$ where the index $\alpha$ accounts for possible degeneracies in the energy spectrum.
The initial energy is measured yielding a value  $E^{\rm in}_i$ with probability $p^{\rm in}_i = \sum_\alpha \bra {e^{\rm in}_{i\alpha}} \rho^{\rm in}\ket {e^{\rm in}_{i\alpha}}$ and leaving the system in the projected state $\rho^{P}_i = \sum_\alpha \ket {e^{\rm in}_{i\alpha}} \bra {e^{\rm in}_{i\alpha}}\rho^{\rm in}\ket {e^{\rm in}_{i\alpha}}\bra {e^{\rm in}_{i\alpha}} $.
We now change the Hamiltonian from $H_{\rm in}$ to $H_{\rm fin}$ in time while the system is isolated from any environment. We denote the eigenvalues and orthonormal eigenvectors of the final Hamiltonian $H_{\rm fin}$  $E^{\rm fin}_i$ and  $\ket {e^{\rm fin}_{i\alpha}}$.  This change induces a unitary evolution so that the state at the end of the process is $U\rho^{P}_i U^\dagger$ where $U$ is the evolution operator. The final energy is measured yielding the value $E^{\rm fin}_j$ with conditional probability $p(j|i) = \sum_\beta \bra {e^{\rm fin}_{j\beta}} U\rho^{P}_i U^\dagger \ket {e^{\rm fin}_{j\beta}}$. The work done on the system for this particular combination of measurement outcomes is $W=E^{\rm fin}_j-E^{\rm in}_i$. The mean work can be obtained by averaging over all possible measurement outcomes:
\begin{equation}
\langle W\rangle = \sum_{ij} p^{\rm in}_i p(j|i) (E^{\rm fin}_j-E^{\rm in}_i)\;.
\end{equation}
 This expression can also be cast as an energy balance:
 \begin{equation}
\langle W\rangle ={\rm Tr}\left(U\rho^PU^\dagger H_{\rm fin}\right) -{\rm Tr}\left(\rho^{\rm in} H_{\rm in}\right), 
\end{equation}
where we have set $\rho^P=\sum_i p^{\rm in}_i\rho^{P}_i$.
If the initial state $\rho^{\rm in}$ is diagonal in the initial energy eigenbasis, i.e. $\rho^{\rm in}=\rho^P$, the average work is just the energy balance of the initial and final states calculated with their respective Hamiltonians. 
Thus, energy is extracted (work is produced) if $W<0$.
In a similar way one can estimate the heat exchanged with the environment during the cooling and heating strokes by replacing the evolution operator with the completely positive map describing the process. 
We thus denote by $-W_1$ and $-W_2$ the work extracted during the compression and expansion strokes, respectively. Similarly, we denote by $Q_1$ and $Q_2$ the heat exchanged with the environments during the heating and cooling, respectively. We assume that $Q>0$ when heat is absorbed from the reservoirs increasing the system's energy. 

Since the evolution is cyclic, we have
\begin{equation}
W_1+W_2+Q_1+Q_2=0\;,
\end{equation}
and thus the total work extracted is
\begin{equation}
W_T=-(W_1+W_2) =Q_1+Q_2\;.
\end{equation}
We define the efficiency in the usual way as the ratio of the work extracted (if positive) and the total heat absorbed by the reservoirs:
\begin{equation}
\eta=\frac{W_T}{\sum_{Q_i>0} Q_i}\;.
\end{equation}

We remark here that the two-time energy measurement protocol may affect both the work produced (see Ref.~\cite{PhysRevLett.118.050601}) and the quantum correlations of the working substance which is the main focus of this paper. In our analysis we present both cases in which an energy observation is performed at each stroke and cases in which it is not.

\section{Engine with local baths}\label{Engine with local baths}
We now start our analysis of the two-qubit engine performance in the case of local baths. To assess this we will use an XX Hamiltonian \cite{1402-4896-88-6-065008}, 
namely we set $J_x=J_y=J$ in Eq.~\eqref{hamiltonian}.
During the heating and cooling strokes each of the two qubits is coupled to a local thermal reservoir. The corresponding dynamics is given by Eq.~\eqref{eq:Masterequation} with Lindblad operators $\mathcal{L}_{\sigma_+^j}$ with coefficient $g_+^j=\gamma_j\bar n_j$ and $\mathcal{L}_{\sigma_-^j}$ with coefficient $g_-^j=\gamma_j(\bar n_j+1)$, where the index $j=a,b$ refers to the two qubits and the jump operators are the usual rising and lowering operators $\sigma_+=|1\rangle\langle0|\;,\;\sigma_-=|0\rangle\langle1|$. The coefficients $\gamma_j$ set the interaction rate of each qubit with its environment and $\bar n_j $ set the corresponding equilibrium temperature $\bar n_j=(e^{2 B_j/T_j}-1)^{-1}$ (setting the Boltzmann constant to 1), where $B_j$ and $T_j$ are the magnetic field and temperature for the bath coupled to qubit $j$, respectively. 

In this generalised Otto cycle, during the compression (expansion) stroke the magnetic field is changed, for both qubits, from $B_1$ to $B_2$ ($B_2$ to $B_1$) with $B_2>B_1$, while the qubits coupling $J$ is held constant. We note that for $B=J$ there is a ``level crossing" in the energy levels. However since the eigenvectors of the XX Hamiltonian are parameter independent the crossing does not affect the qubits' evolution. Note that, in the presence of inter-qubit coupling, the heating and cooling steady states of each qubit are generally not thermal states, not even for equal reservoirs temperatures, the difference significantly increasing with their mutual interaction. In other words, the system under scrutiny is out of equilibrium.
In Sec.~\ref{Differences between steady state and thermal state} we make a detailed comparison between the steady state of Eq.~\eqref{eq:Masterequation} and the corresponding equilibrium thermal state.

\subsection{Equal temperatures}\label{Loc_A}

In the case of equal temperatures $\bar n_j=\bar n$ and for equal rates $\gamma_j=\gamma$, the steady state is diagonal in the $\sigma_{zj}$ eigenbasis (given by the set ${\ket{00}, \ket{01}, \ket{10}, \ket{11}}$):
\begin{equation}\label{steadystate}
\rho_{S}=\frac 1Z
\begin{pmatrix}
\bar{n}^2& 0 & 0 &0 \\
0& \bar{n}(1+\bar{n}) & 0 & 0 \\
0& 0 & \bar{n}(1+\bar{n}) & 0\\
0& 0 & 0 &(1+\bar{n})^2
\end{pmatrix}
\end{equation}
with $Z=(1+2\bar{n})^{2}$  a normalisation constant. The latter state, being a product state, shows no quantum correlations, not even in the form of quantum discord. In addition, due to the symmetries of the model, the steady state $\rho_S$ is invariant under an energy measurement, implying that the projected state $\rho^P$ coincides with $\rho_S$. As a consequence, for an engine operating under these conditions, there will be no effects at all related to quantum correlations. This can nevertheless be used as a benchmark to compare to other cycles.  
 
 Assuming that the heating process corresponds to connecting the two qubits with two local baths with thermal occupation $\bar n_H$ and, similarly, the cooling process with $\bar n_C$ with $\bar n_C<\bar n_H$, one can obtain the following values for the work and heat contributions:
 \begin{eqnarray}\label{localbathheatsandwork}
W_1&=&\dfrac{2(B_1-B_2)}{1+2\bar{n}_C} \;, \\
Q_1&=& 2B_2\left(\dfrac{1}{1+2\bar{n}_C}-\dfrac{1}{1+2\bar{n}_H}\right) \;,\\
W_2&=& -\dfrac{2(B_1-B_2)}{1+2\bar{n}_H} \;,\\
Q_2&=& 2B_1\left(\dfrac{1}{1+2\bar{n}_H}-\dfrac{1}{1+2\bar{n}_C}\right) \;.
\end{eqnarray}
Note that the thermodynamic quantities above do not depend on the coupling $J$ and are actually twice as large as the corresponding values of a single qubit Otto engine. The total work is 
\begin{equation}\label{enegineperfomance}
W_T= \dfrac{4(B_1-B_2)(\bar{n}_C-\bar{n}_H)}{(1+2\bar{n}_C)(1+2\bar{n}_H)}\;,
\end{equation}
and the efficiency turns out to be given by the standard Otto efficiency $\eta= 1-B_1/B_2$,
 which is based solely on the ratio between the two values of $B$ used. 
 
 In Appendix~\ref{app:partial} we provide an analysis of the engine assuming only partial thermalisation during the cooling and heating strokes.
  As mentioned earlier, this setting provides a specific example in which the working substance displays no quantum correlations, which therefore cannot have any relation with the thermodynamic quantities that characterise the engine.

 \subsection{Different qubits temperatures} \label{Loc_B}
We now consider the more general situation in which the temperatures of the cooling and heating baths coupled to the two qubits are different. We call $\bar n_{Ca}$ and $\bar n_{Cb}$ the corresponding cooling bath populations and, similarly, we define $\bar n_{Ha}$ and $\bar n_{Hb}$ for the heating stroke. For the sake of generality, we also assume different decay rates $\gamma_a$ and $\gamma_b$.

Under this condition, we can see that the steady state is no longer diagonal, as in Eq.~\eqref{steadystate}, but is of the form:
\begin{equation}
\label{unequalsteadystate}
\rho_S=\frac{1}{\alpha}\begin{pmatrix}
r_{11}& 0 & 0 &0 \\
0& r_{22} &i r_{23} & 0 \\
0& -ir_{23}  &r_{33} & 0\\
0& 0 & 0 &r_{44}
\end{pmatrix}\;,
\end{equation}     
where $\alpha$ is a normalisation constant. The analytical expressions of $\alpha$ and all the real coefficients $r_{ij}$ are given in Appendix \ref{app1}.

Depending on the inter-qubit couplings, the reservoir temperatures, and the decay rates, the steady state might become entangled. To measure the amount of the latter we employ the concurrence \cite{PhysRevLett.80.2245}, defined as follows. Let $\sigma$ be the density matrix of two qubits. We define the matrix $\tilde\sigma=(\sigma_{ya}\otimes\sigma_{yb})\sigma^*(\sigma_{ya}\otimes\sigma_{yb})$ and $\lambda_i$ as the eigenvalues of the positive semidefinite matrix $\sigma\tilde\sigma$ sorted in decreasing order. Then the concurrence of $\sigma$ is defined as:
\begin{equation}
C(\sigma)=\max\left(0,\sqrt{\lambda_1}-\sqrt{\lambda_2}-\sqrt{\lambda_3}-\sqrt{\lambda_4}\right).
\end{equation}

\begin{figure}
 	\centering
 	\includegraphics[height=3cm]{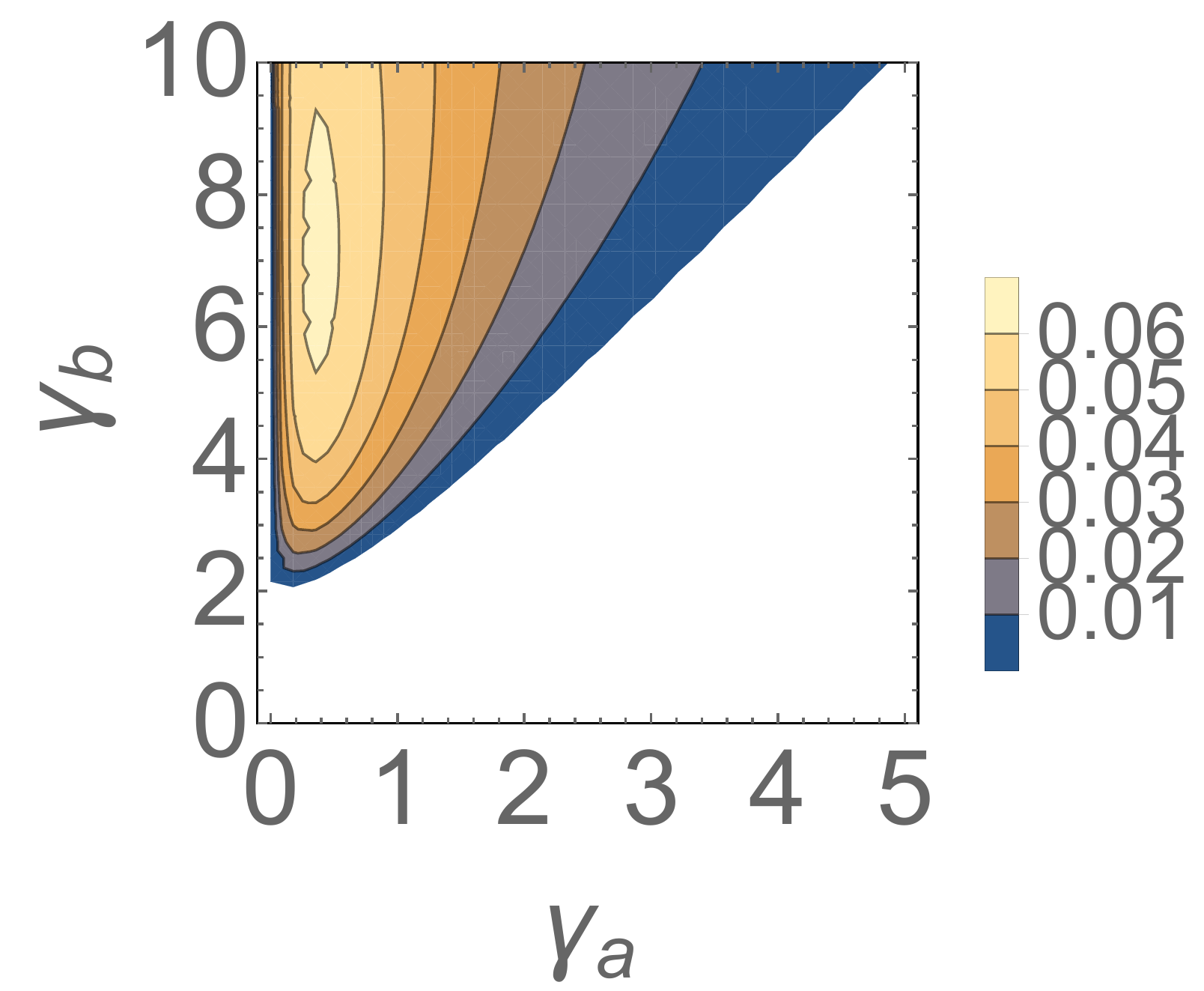}
 	\includegraphics[height=3cm]{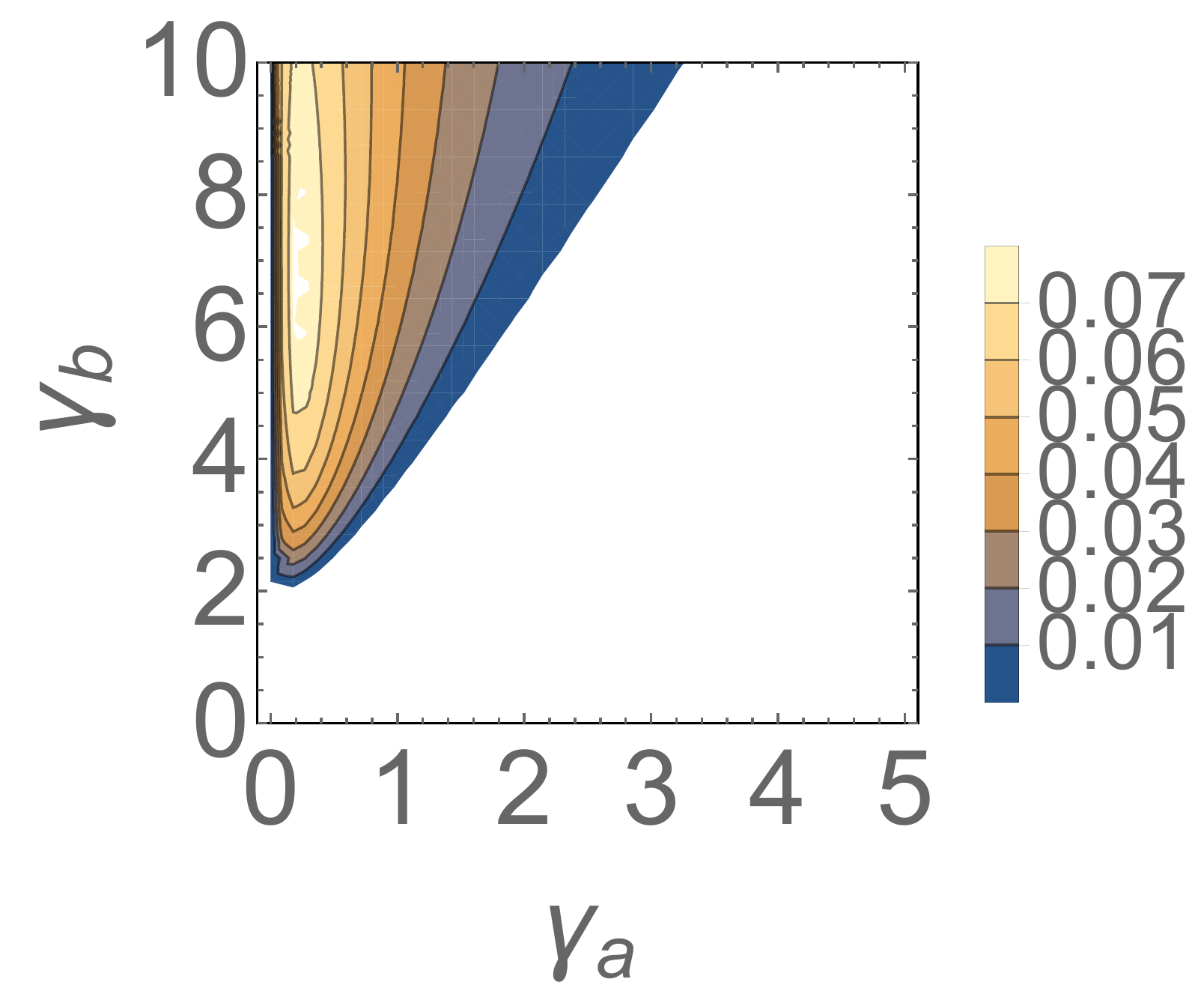}
 	\caption{Contour plot of the concurrence of state $\rho_S$ in Eq.~\eqref{unequalsteadystate} corresponding to $\bar{n}_{b}=0$ and $\bar{n}_a=1$ (left) and $\bar{n}_a=2$ (right) as a function of $\gamma_a$ and $\gamma_b$, with $J=1$ for both panels.}
 	\label{ConcurranceVsgamma}
 \end{figure}

\begin{figure}
	\centering
	\includegraphics[height=4cm]{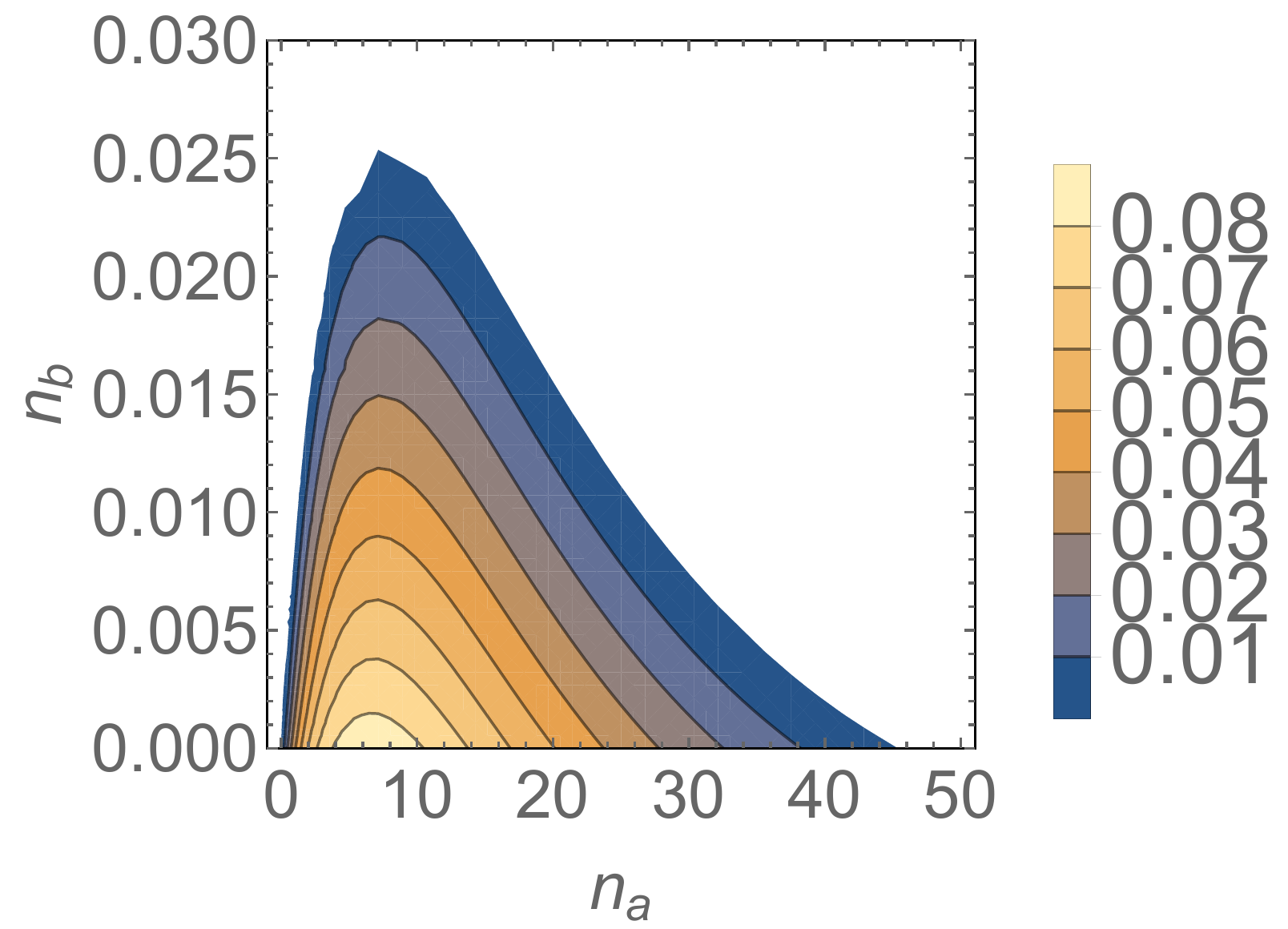}
	\caption{Contour plot of the concurrence of the  state  \eqref{unequalsteadystate}, $\gamma_a=0.1,\gamma_b=5$ $J=1$ as a function of $\bar{n}_a$ and $\bar{n}_b$.  }
	\label{ConcurranceVsnbar}
\end{figure}

State $\rho_S$ in Eq.~\eqref{unequalsteadystate} is entangled for certain values of $\gamma_a$ and $\gamma_b$, as shown in Fig.~\ref{ConcurranceVsgamma} for the extreme case in which $\bar{n}_{ib}=0$, $i=\{C,H\}$. Not surprisingly, we observe that the region of parameters in which the state is entangled shrinks as the reservoir temperatures increases. We see that, in order for the working substance to sustain entanglement, not only do the bath temperatures but also their rates have to be unequal. Note also that, even at its maximum level, the amount of entanglement is small. Lifting the restriction $\bar{n}_{ib}=0$ ($i=\{C,H\}$), we can see in Fig.~\ref{ConcurranceVsnbar} that the entanglement persists only for very low values of $\bar{n}_{ib}$.

Having found the presence of entanglement in the working substance at steady state, we now look at its possible effect on the performance of the Otto engine. 
Contrary to the case in Sec.~\ref{Loc_A}, the projected state after the energy measurement is now different from $\rho_S$:
\begin{equation}
\rho^P=
\begin{pmatrix}
r_{11}& 0 & 0 &0 \\
0& \frac{r_{22}+r_{33}}{2} &0 & 0 \\
0& 0  &\frac{r_{22}+r_{33}}{2}& 0\\
0& 0 & 0 &r_{44}
\end{pmatrix}\;.
\end{equation}     
In particular, due to the symmetries of the system, $\rho_P$ is diagonal and therefore it does not bear any entanglement or quantum correlations. Note however that the measurement process has no effect on the elements $r_{11}$ and $r_{44}$ which, in turn, solely determine the energy:
\begin{equation}\label{XXEnergy}
\langle H\rangle = B(r_{11}-r_{44})\;.
\end{equation}
Therefore, despite destroying all quantum correlations, as a point of fact, the energy measurement process has no effect on the energetic performance of the engine. In particular, the analytical expression for the total amount of work is very long and is reported in Appendix \ref{app1} for $\bar{n}_{ib}=0$ ($i=\{C,H\}$). It is possible to prove in general that for this model the efficiency is the same as the Otto cycle.
In fact, let us call $\rho^C$ and $\rho^H$, with elements $r_{ij}^H$ and $r_{ij}^C$, the steady states of the cold and hot bath, respectively. 
Then the total work reads:
\begin{equation}\label{unequalwork}
W_T
=(B_2-B_1)(r_{11}^H-r_{11}^C+r_{44}^C-r_{44}^H)
\end{equation}
while the heat absorbed from the two hot reservoirs is
\begin{equation}
Q_1
=B_2(r_{11}^H-r_{11}^C+r_{44}^C-r_{44}^H)
\end{equation}
and thus
\begin{equation}\label{unequaleta}
\begin{matrix}
\eta =\dfrac{W_T}{Q_1} = 1-\dfrac{B_1}{B_2}.
\end{matrix}
\end{equation}
We see that the efficiency of the engine is related exclusively to the values of $B$ that are used in the cycle, which are in turn unrelated to the level of entanglement present in the system before the measurement process. In addition, the work of the system is only dependent on the populations of the system and is thus again unaffected by the entanglement.

Summarising, this setting provides an example in which the working substance does display quantum correlations in the form of entanglement. However, no relation exists between the latter and the thermodynamic quantities that characterise the engine. In fact, on one hand, those quantum correlations are destroyed during the measurement process and, on the other hand, the design of this engine itself implies that its energetic performances depend on the populations only, which are left untouched by the measurement.

We briefly mention that the dissipative evolution of two qubits interacting via the XX Hamiltonian and described by a local master equation was investigated in \cite{BaraReaptedinteaction}. Utilising their calculations, it is straightforward to see that the external work required to generate such dissipative evolution for our two-qubit system is 0 since the magnetic field is identical on both spins.

\subsection{Differences between steady state and thermal state } \label{Differences between steady state and thermal state}
Our previous discussion has exclusively focused on the dynamics implied by the master equation ,\eqref{eq:Masterequation}, and, in particular, its associated steady states. The latter are crucial in determining the performances of the engine, and it is therefore important to consider carefully their emergence. 

When connecting a system to an equilibrium reservoir, one would expect for long times the emergence of a thermal state: $\rho^T(H)=e^{-\beta H}/\tr[e^{-\beta H}]$. Note that the thermal state assumes that the two qubits interact with equal temperature baths. For the XX Hamiltonian the thermal state is given by
\begin{equation}\label{thermal state}
\rho^T(H)=\frac{1}{Z}
\begin{pmatrix}
e^{-\beta B} &0 & 0 &0 \\
0& \cosh(J\beta) & -\sinh(J\beta) & 0 \\
0&-\sinh(J\beta) & \cosh(J\beta)  & 0\\
0& 0 & 0 &e^{\beta B} 
\end{pmatrix}
\end{equation}
with $Z=e^{-\beta B}+e^{\beta B}+e^{-\beta J}+e^{-\beta J} $ being the partition function. We see that the state in Eq.~\eqref{thermal state} is different from that in  Eq.~\eqref{steadystate}, notably the thermal state is not diagonal and can be entangled if $\sinh^2J\beta>1$. Thermodynamics engines in which the working substance operates between thermal states have been considered for example in Ref.~\cite{PhysRevE.90.032102, PhysRevA.96.052119, PhysRevE.83.031135, PhysRevE.92.022142, 1402-4896-88-6-065008}.

\begin{figure}
	\centering
	\includegraphics[height=2.8cm]{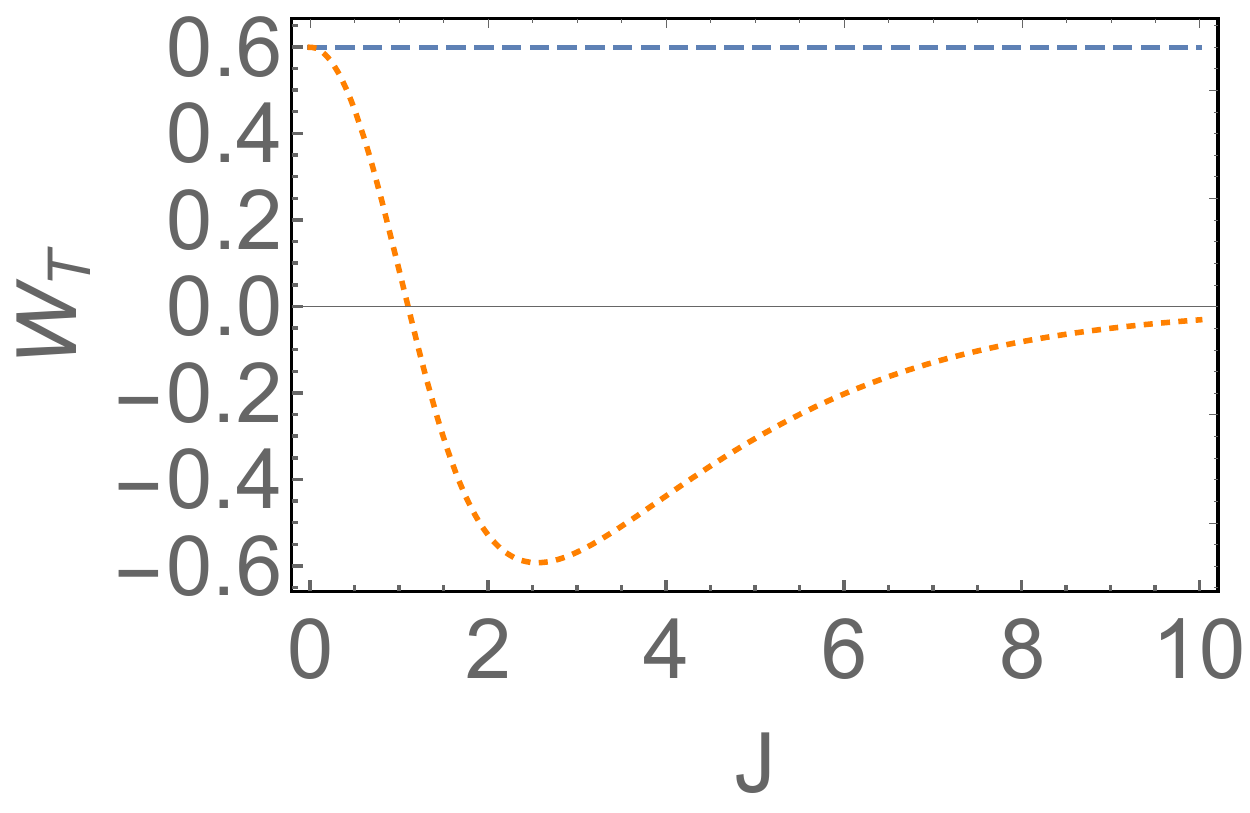}
	\includegraphics[height=2.8cm]{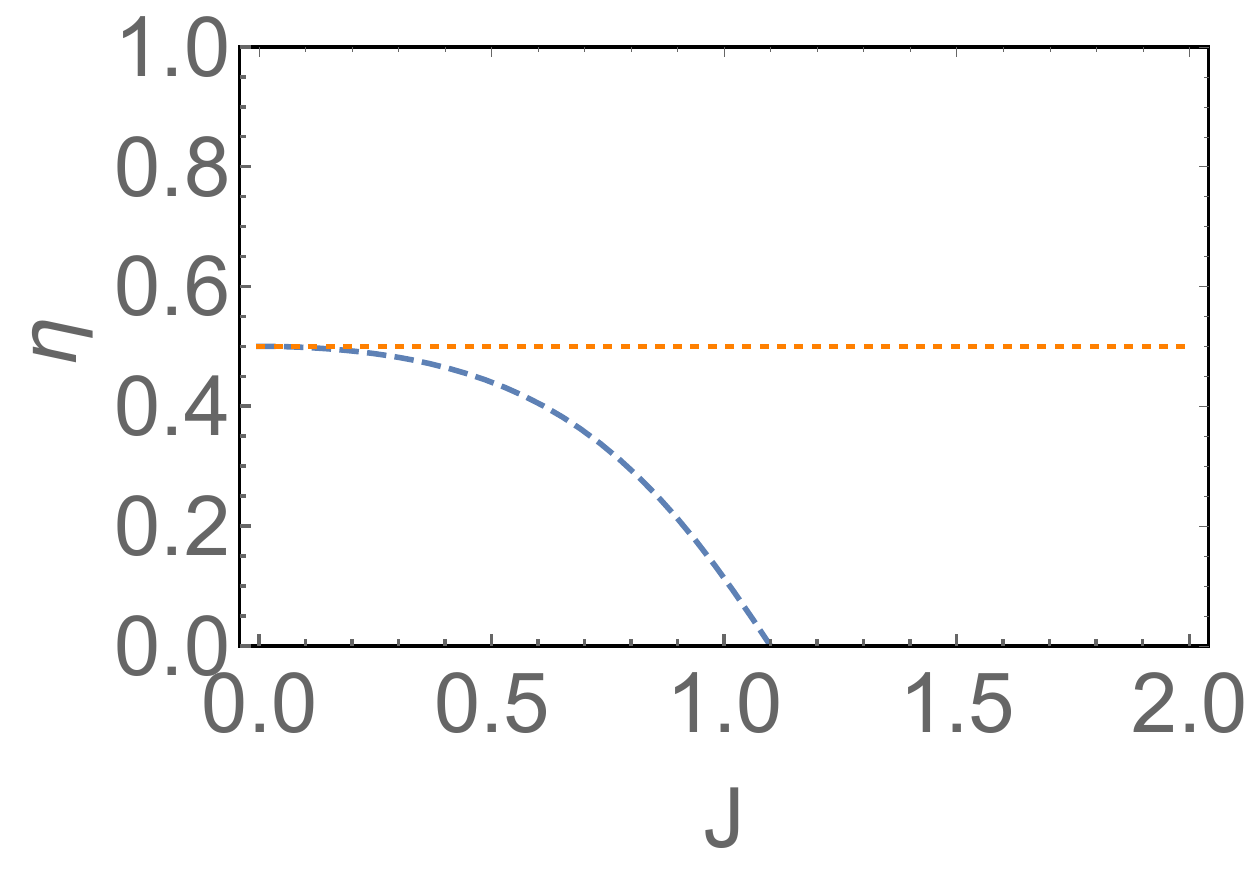}
	\includegraphics[height=2.8cm]{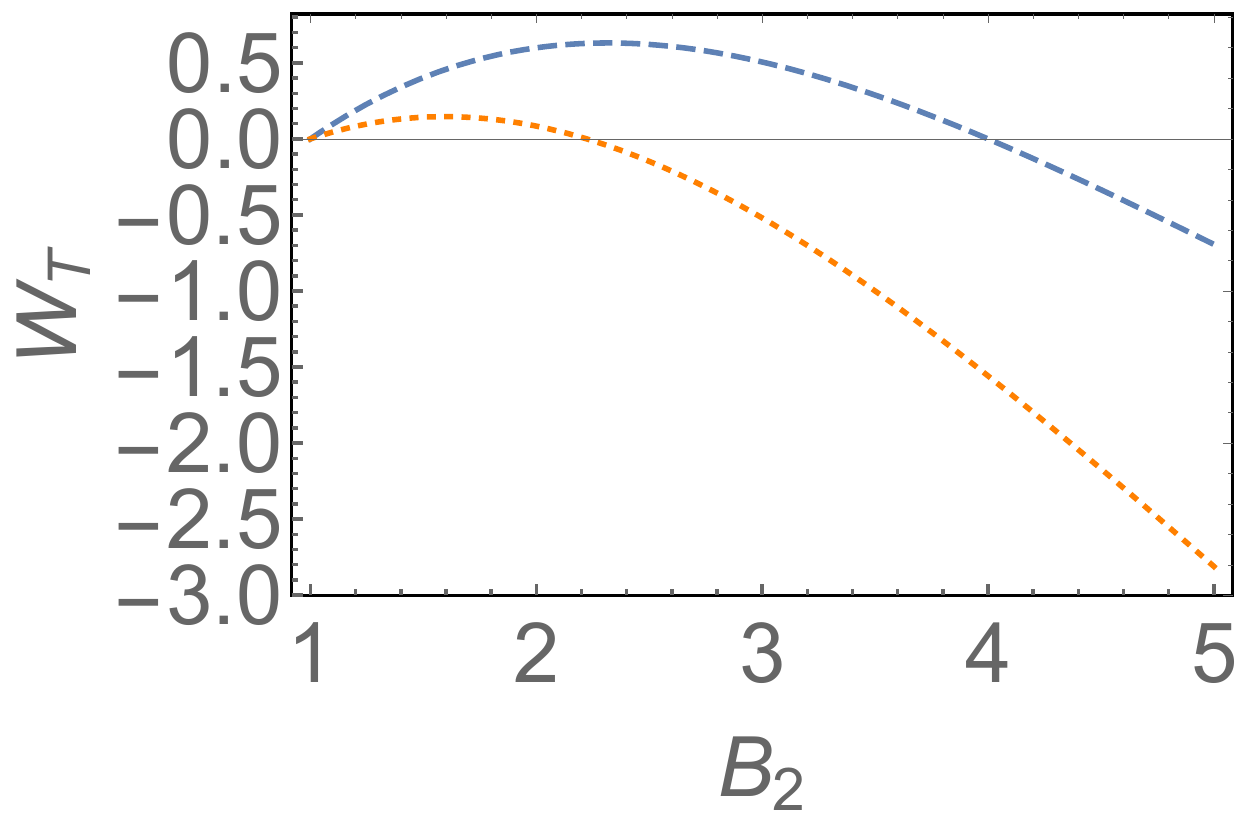}
	\includegraphics[height=2.8cm]{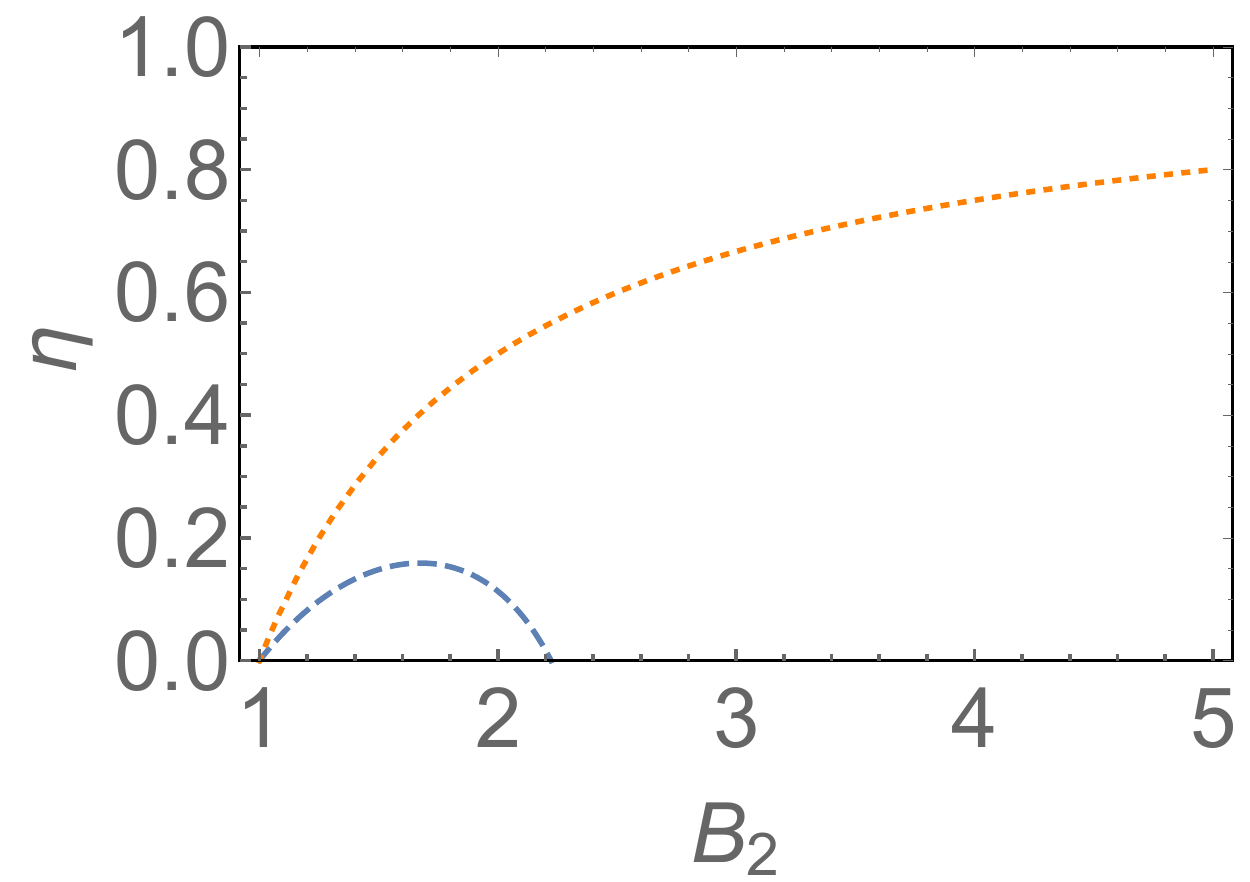}
	\caption{Plots of the work and efficiency for an Otto cycle operating between $T_1=1$ and $T_2=4$ using the steady state (dotted) and the thermal state (dashed). Top: We fix $B_1=1$ and $B_2=2$ and in the compression and expansion strokes we change $J$ from 1 to the value reported on the horizontal axis. Bottom: We fix $J=1$ and vary the magnetic field from $B_1=1$ to $B_2$ reported on the horizontal axis.}
	\label{ThermalSteadyComparisions}
\end{figure}

We now see that the difference between thermal and steady states implies that the work produced by an engine operating with the former is different from the one given in Eq.~\eqref{enegineperfomance}. Let us start by substituting the definition of $\bar{n}_j$ into Eq.~ \eqref{enegineperfomance}:
\begin{equation}\label{BTwork}
W_T= 2(B_1-B_2)\left[\tanh\left(\dfrac{B_1}{T_1}\right)-\tanh\left(\dfrac{B_2}{T_2}\right)\right]\;,
\end{equation}   
which allows us to compare the two scenarios as a function of temperatures and magnetic fields. Calculating the energetic performance of an engine operating with thermal states we have
 \begin{eqnarray}\label{THermalbathheatsandwork}
W_1&=&2(B_1-B_2)\Phi_1, \\
Q_1&=&\Theta_{1}-\Theta_{2},\\
W_2&=& 2(B_2-B_1)\Phi_2, \\
Q_2&=& \Theta_{2}-\Theta_{1},\\[1.5ex]
W_T&=& 2(B_1-B_2)(\Phi_2 -\Phi_1) \label{enegineperfomance_thermal},\\[1.5ex]
\eta&=&\dfrac{2(B_1-B_2)(\Phi_2 -\Phi_1)}{\Theta_{1}-\Theta_{2}},
\end{eqnarray}
where we have defined $\Theta$ and $\Phi$ as 
\begin{eqnarray}
\Theta_i&=&\dfrac{\sinh\left(\frac{2B_i}{T_i}\right)+J\sinh\left(\frac{2J}{Ti}\right)}{\cosh\left(\frac{2B_i}{T_i}\right)+\cosh\left(\frac{2J}{T_i}\right)}\label{Theta},\\
\Phi_i&=&\dfrac{\sinh\left(\frac{2B_i}{T_i}\right)}{\cosh\left(\frac{2B_i}{T_i}\right)+\cosh\left(\frac{2J}{T_i}\right)}\label{Phi}.
\end{eqnarray}
First, note that when $J$ is 0 the work done is identical to  \eqref{BTwork}. When $J\neq0$, though,  the two scenarios diverge significantly as the thermal-state work in Eq.~\eqref{enegineperfomance_thermal} depends on $J$ while the steady-state work in Eq.~\eqref{enegineperfomance} does not. A plot of the differences between these two scenarios is given in Fig. \ref{ThermalSteadyComparisions}, where the efficiency of the cycle is also shown.

The calculations above show that care must be taken when comparing models of quantum engines with a working medium formed by several interacting particles. In particular, different relaxation processes the particles are subject to can lead to different steady states, which, in turn, are  determined not solely by the effective bath temperatures but also by the underlying physics. Working substances whose open dynamics can be legitimately described by local baths can in fact determine engine performances strikingly different from the ones associated with engines with working substances operating between non-local (or common) baths. In the next section we  further elucidate the relevance of common baths. In particular, we present settings in which, contrary to what is found in Secs.~\ref{Loc_A} and \ref{Loc_B}, a clear relation between the engine performance and quantum correlations can be drawn.

\section{Common Baths and Engine Performance}\label{Common Baths and Engine Performance.}
From the analysis above we see that local baths have limited possibility in the generation of entanglement, in that the amount of the latter is usually small and can be destroyed during the measurement process. In fact, we have provided two settings, with different features from the correlation viewpoint, for which ,nonetheless, no relation exists between quantum correlations and thermodynamics performances. In this section, we generalise the settings under scrutiny by considering common Lindblad quantum jump operators acting on the working substance. This allows for  greater freedom in generating coherences and quantum correlated steady states. In particular, we examine a range of Hamiltonians of the form given in Eq.~\eqref{hamiltonian} and see how the average energy $ \langle H \rangle $ is affected by coherences at steady state. We then use this model to design a generalised Otto cycle heat engine whose performance is related to the quantum correlations in the working substance. We stress, once more, that the reservoirs we are considering are not thermal baths and therefore do not necessarily bring the system to an equilibrium state and might require external work to operate. Nevertheless, they are physical models that can be engineered experimentally.

\subsection{Non-interacting Hamiltonian}\label{Local Hamiltonian}
We begin our analysis by looking at non-interacting Hamiltonians, setting $J_x=J_y=0 $ in Eq.~\eqref{hamiltonian}. In this setup there can be no correlations ascribable to the Hamiltonian dynamics, so any entanglement in the system will be generated by a carefully chosen common environment. We start with a Hamiltonian, similar to Eq.~\eqref{hamiltonian} but a with magnetic field aligned along the $x$ axis:
\begin{equation}
H=B (\sigma_{xa}+\sigma_{xb}).
\end{equation}
 The corresponding mean energy depends on the entries $r_{ij}$ of the density matrix as:
\begin{equation}\label{Xenergy}
\langle H \rangle=2B\;{\rm Re}(r_{12}+r_{13}+r_{24}+r_{34}),
\end{equation}
namely, on the real part of the coherences of the system's state. 

We assume each qubit to be coupled to its local thermal bath at the same temperature, in terms of $\bar n$, and described by the following jump operators and strengths:
\begin{equation}
\label{eq:localdissipation}
\begin{matrix*}[l]
a_1 = \sigma_+^a,\quad & g_1=(1-\gamma)\bar n \;;\\
a_2 =\sigma_-^a,\quad& g_2=(1-\gamma)(\bar n+1)\;;\\
a_3={\sigma_+^b},\quad & g_3=(1-\gamma)\bar n \;;\\
a_4={\sigma_-^b},\quad& g_4=(1-\gamma)(\bar n+1).
\end{matrix*}
\end{equation}
Additionally, the qubits are coupled to a common environment described by two jump operators:
\begin{equation}
\begin{matrix*}[l]
a_5 = {(\sigma_-^a-\sigma_z^b)(\sigma_-^a-\sigma_-^b)},\quad & g_5=\gamma\;, \\
a_6 ={(\sigma_z^a-\sigma_x^b)(\sigma_x^a-\sigma_-^b)},\quad & g_6=\gamma\;.
\end{matrix*}
\end{equation}
The parameter $\gamma$ measures the relative strength of the common bath compared to the local baths. The common baths above were inspired by the bi-local jump operators for two qubits as described in \cite{0295-5075-116-1-14002} but adapted to maximise \eqref{Xenergy} and thus the coherences in the system. The steady state of the corresponding master equation can be found analytically but its expression is long and we do not report it.

With this setup we design a non-equilibrium quantum engine. As before during the compression (expansion) strokes the magnetic field is changed from $B_1$ to $B_2$ ($B_2$ to $B_1$) with $B_2>B_1$. For the cooling stroke, we consider the environment described by the jump operators $a_i$ with strengths $g_i$ with $\gamma =0$. For the heating stroke we take a finite value $\gamma\ne0$. The temperature of the local thermal baths described by $\bar n$ is kept constant. Note that, as mentioned above, the nomenclature used here is related to the fact that during the cooling (heating) stroke the average system energy decreases (increases).

\begin{figure}
	\centering
	\includegraphics[height=5cm]{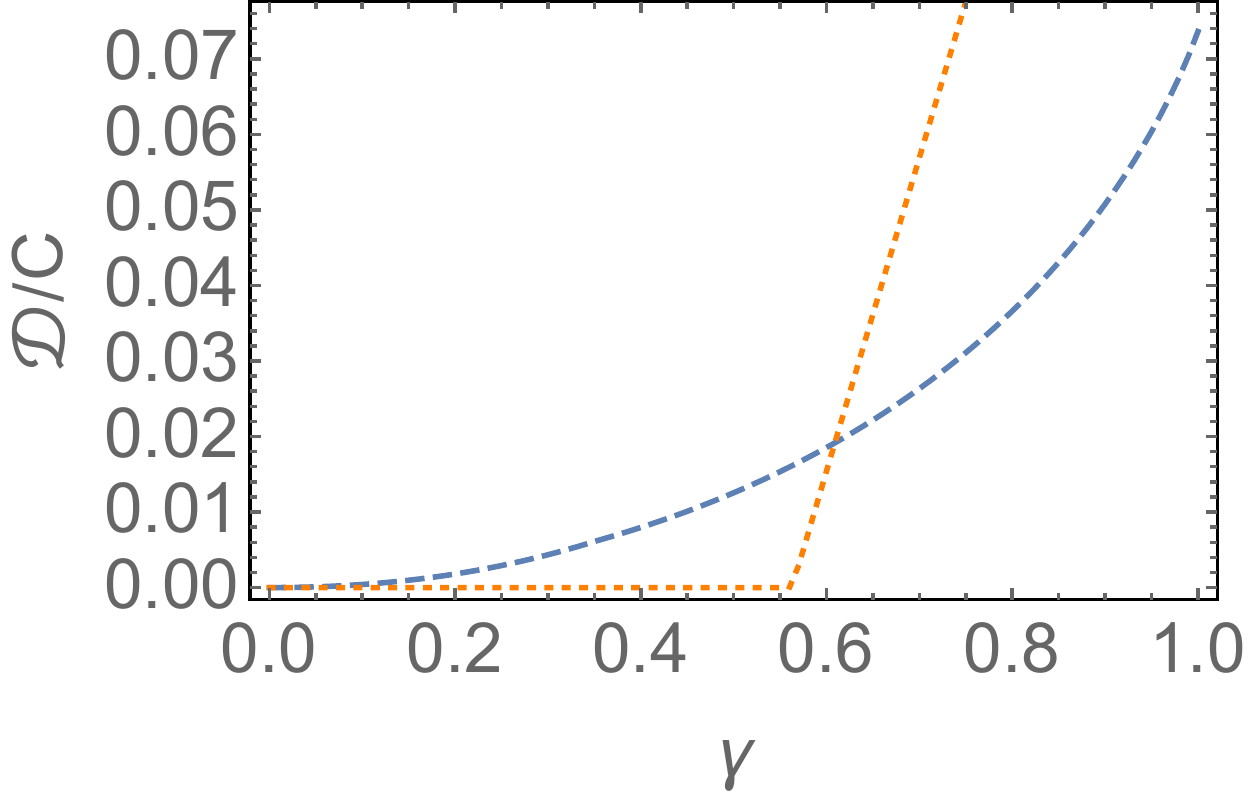}
	\includegraphics[height=5cm]{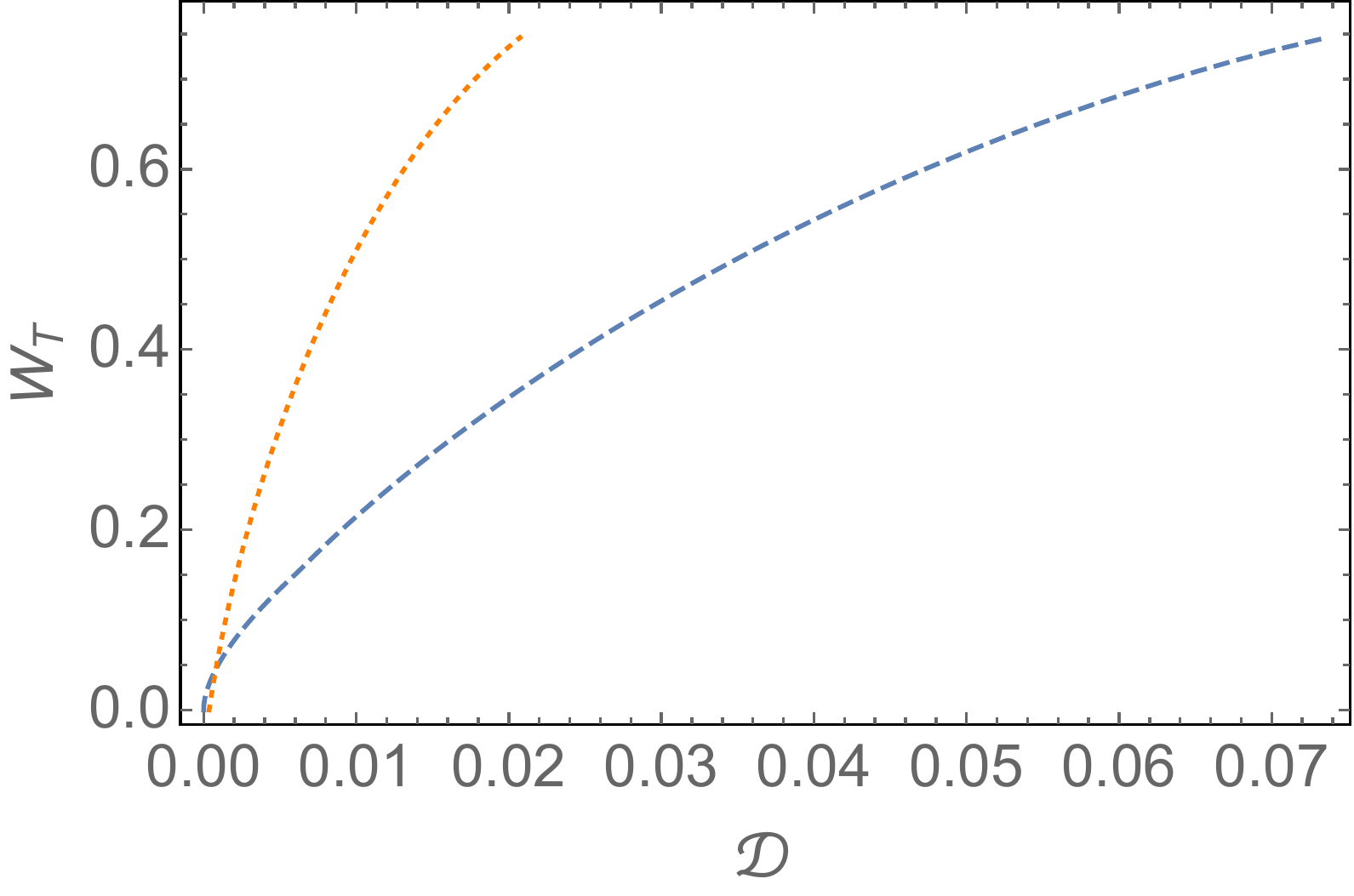}
	\caption{Common baths with non-interacting Hamiltonians. (Top) Plot of the discord (dashed curve) and concurrence (dotted curve ) of the steady state of the common bath introduced in Sec.~\ref{Local Hamiltonian} for values of $\gamma$ from 0 to 1, $\bar{n}=1,B_1=1,B_2=2$. (Bottom) Total work produced as a function of the discord of the steady state of the hot bath for $0<\gamma<1$ before measurement is performed (dotted curve) and after measurement (dashed curve). Same parameters as in the top panel.}
	\label{XFieldworkVsDiscord}
\end{figure}

Considering the steady states that correspond to such an engine, one can show that the coherences are not large enough to generate entanglement for any value of $\gamma$ (see Fig.~\ref{XFieldworkVsDiscord}). Hence, rather than the concurrence, in order to measure the quantum correlations between the two qubits we use the quantum discord $\mathcal{D}(\rho_S)$ of the steady state, whose definition is reported in Appendix \ref{app:discord}. We can see in Fig.~\ref{XFieldworkVsDiscord} that discord is always present in the system as long as $\gamma\neq 0$, even when the concurrence is absent. This is actually the case for typical multipartite open systems \cite{ferraro2010almost}.

We calculate the total work done by the engine and the discord of the steady state after connecting to the hot bath for values of $\gamma$ between 0 and 1.
The results in Fig.~\ref{XFieldworkVsDiscord} show that the discord of the hot steady state increases as the total work done by the system increases. This shows a clear relation, for this type of engine, between quantum correlations and energetic performances. The crucial features that determine such a relation are, on the one hand, the dependence of the energy on the coherences present in the working substance [Eq.~\eqref{Xenergy}] and, on the other hand, the presence of a common bath. Note that this relation remains valid even when  the energy measurement related to the two-time definition of work takes place. The effect of the measurement, as shown in Fig.~\ref{XFieldworkVsDiscord}, is to decrease the amount of discord which however remains non-zero.

\subsection{Commom baths and interacting Hamiltonians}\label{Common_interacting}
We now consider the effect of common baths on the functioning of an engine comprised of two interacting qubits, with an Ising Hamiltonian, \eqref{hamiltonian}, with $B=0$ and $J_x=J,J_y=0$. As $J$ is the only parameter present in the Hamiltonian it will be the one that is raised (lowered) during the compression (expansion) strokes between two values $J_1<J_2$. This can be realised, for example, in quantum simulation experiments with trapped ions~\cite{friedenauer2008simulating}. The average energy in terms of the state density matrix elements and for a coupling $J$ reads
\begin{equation}\label{XInteraction}
\langle H \rangle=2J\;{\rm Re}(r_{14}+r_{23}).
\end{equation}
 As in the previous case, we see a dependence of the average energy on the coherences of the system. We want to construct a heat cycle in which the work and heat exchanged depend on these off-diagonal entries. To this aim, we  consider two scenarios described below. 
 
{\it Common dephasing and Bell pumping.-}
In the first scenario we assume the working medium to be simultaneously coupled to two common environments. The first is a common dephasing reservoir as found in
Ref.~\cite{0295-5075-101-6-60005} and modelled by the following jump operators and strengths:
 \begin{equation}\label{dephasingbath}
 \begin{matrix}
 &a_1=\sigma_{za}+\sigma_{zb},\quad g_1=(1-\gamma) \;;\\
 &a_2=\sigma_{za}-\sigma_{zb},\quad g_2=(1-\gamma) \;.
 \end{matrix}
 \end{equation} 
 The steady state of this bath is a purely diagonal state. 
 
 The other reservoir drives the system to the Bell state $\ket{\psi^-}=2^{-1/2}(\ket{01}-\ket{10})$ and is described by the following jump operators \cite{JOUR}:
 \begin{equation}
 \begin{matrix}\label{mixedbath}
&a_3=\frac{1}{2}\sigma_{xb}(I+\sigma_{za}\sigma_{zb}),\quad g_3=\gamma\;; \\
&a_4=\frac{1}{2}\sigma_{zb}(I+\sigma_{xa}\sigma_{xb}),\quad g_4=\gamma\;
 \end{matrix}
 \end{equation}
where $I$ is the identity matrix for the two qubits.
In this model, $\gamma$ measures the strength of the Bell state bath over the dephasing one. The system is entangled for all $\gamma\ne 0$, whereas for $\gamma=0$ the steady state is $\ket{01}\bra{01}+\ket{10}\bra{10}$, which is separable. For $\gamma=1$ the steady state is the maximally entangled Bell state $\ket{\psi^-}$ ,and for all other values of $\gamma$, the resulting steady state is:
 \begin{equation}\label{XInteractionsteadystate}
\rho_S= \begin{pmatrix}
 0 &0 &0 &0 \\
 0&1/2 &\mu &0 \\
 0&\mu &1/2 &0\\
 0&0 &0 &0
 \end{pmatrix}\;,
 \end{equation}
 where $\mu=\frac{\gamma}{14\gamma-16}$ , $-1/2<\mu<0$ for $0\le\gamma\le 1$.  
We note that the average energy in Eq.~\eqref{XInteraction} of the steady state $\rho_S$ is $\langle H\rangle =2\mu J$. Thus, for $\gamma=1$, the steady state being $\ket{\psi^-}$, the energy is $-J$ which is lower than the energy of the steady state for $\gamma=0$ (for which $\langle H\rangle =0$).
We thus use the environment with $\gamma\neq0$ as a cold bath, since it decreases the system energy, and the environment with $\gamma=0$ as a hot bath, as the system energy is increased. The work extracted during the cycle is given by:
\begin{equation}\label{xIwork}
W_T=\mu(J_1-J_2)\;.
\end{equation}
Crucially, the concurrence of the steady state $\rho_S$ of the cold bath is $C=-2\mu$. Hence we see that, for the case of this engine, the total work is proportional to the concurrence. This establishes a direct link between the entanglement within the working substance and the energetic performances of the engine.

As before, we note that this link between quantum correlations, this time expressed in the strong form of entanglement, and performances persists even when the measurement process takes place. In fact, the energy eigenstates are in this case the four Bell states $\ket{\psi^\pm}=2^{-1/2}(\ket{00}\pm\ket{11}),\ket{\phi^\pm}=2^{-1/2}(\ket{01}\pm\ket{10})$, independently of $J$.  By noting that the steady state in Eq.~\eqref{XInteractionsteadystate} can be written as:
\begin{equation}\label{Commonsteadystate}
\rho_S = \frac{1+2\mu}{2} \ket{\psi^+}\bra{\psi^+}+
\frac{1-2\mu}{2} \ket{\psi^-}\bra{\psi^-}\;,
\end{equation}
one can immediately conclude that the energy measurements leave the steady state invariant, namely, $\rho_S=\rho^P$. As a consequence, the relation between concurrence $C$ and total work $W_T$ holds true also after the measurement related to the two-time work definition.

Note however that also in the case of this engine, as per the previous ones, the efficiency is just given by the Otto efficiency,
\begin{equation}\label{xIefficenicy}
\eta=1-\dfrac{J_1}{J_2}\;,
\end{equation}
where $J$ plays the role of $B$, as that is the parameter that has been changed in the compression and expansion strokes. 
As the efficiency of the engine does not change as $\gamma$ increases the increase in the work output causes a corresponding increase in the heat input.

 {\it Local dissipation and Bell pumping.-}
The preceding example is a specific case whose construction simplifies many aspects. However, in order to include possible thermal effects, we now consider an environment characterised by the set of jump operators defined in Eq.~\eqref{mixedbath} jointly with the set 
of local thermal dissipators defined in \eqref{eq:localdissipation}.
The steady state for an arbitrary $\gamma$ is a mixture of $\ket{\psi^-}$ and the steady state of the master equation in the absence of common environment, see Eq.~\eqref{steadystate}.

\begin{figure}
	\centering
	\includegraphics[height=5cm]{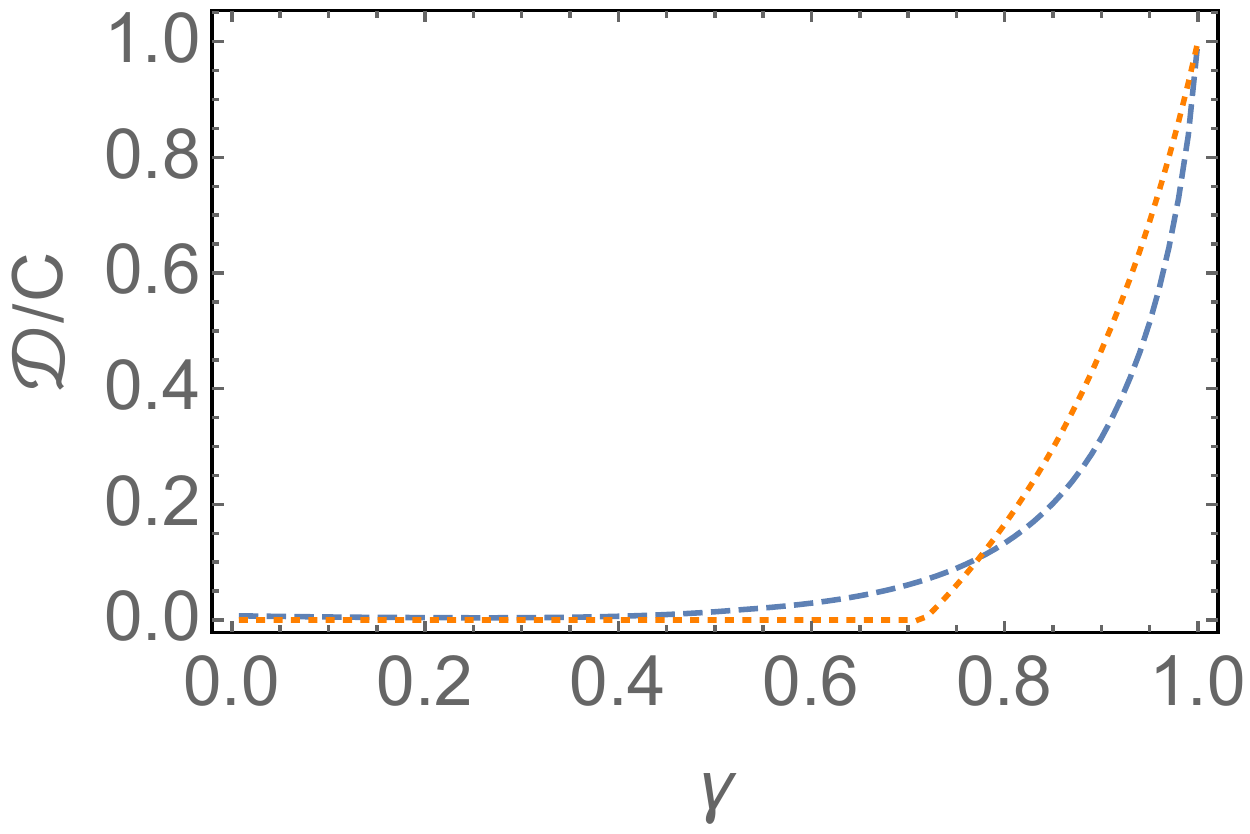}
	\includegraphics[height=5cm]{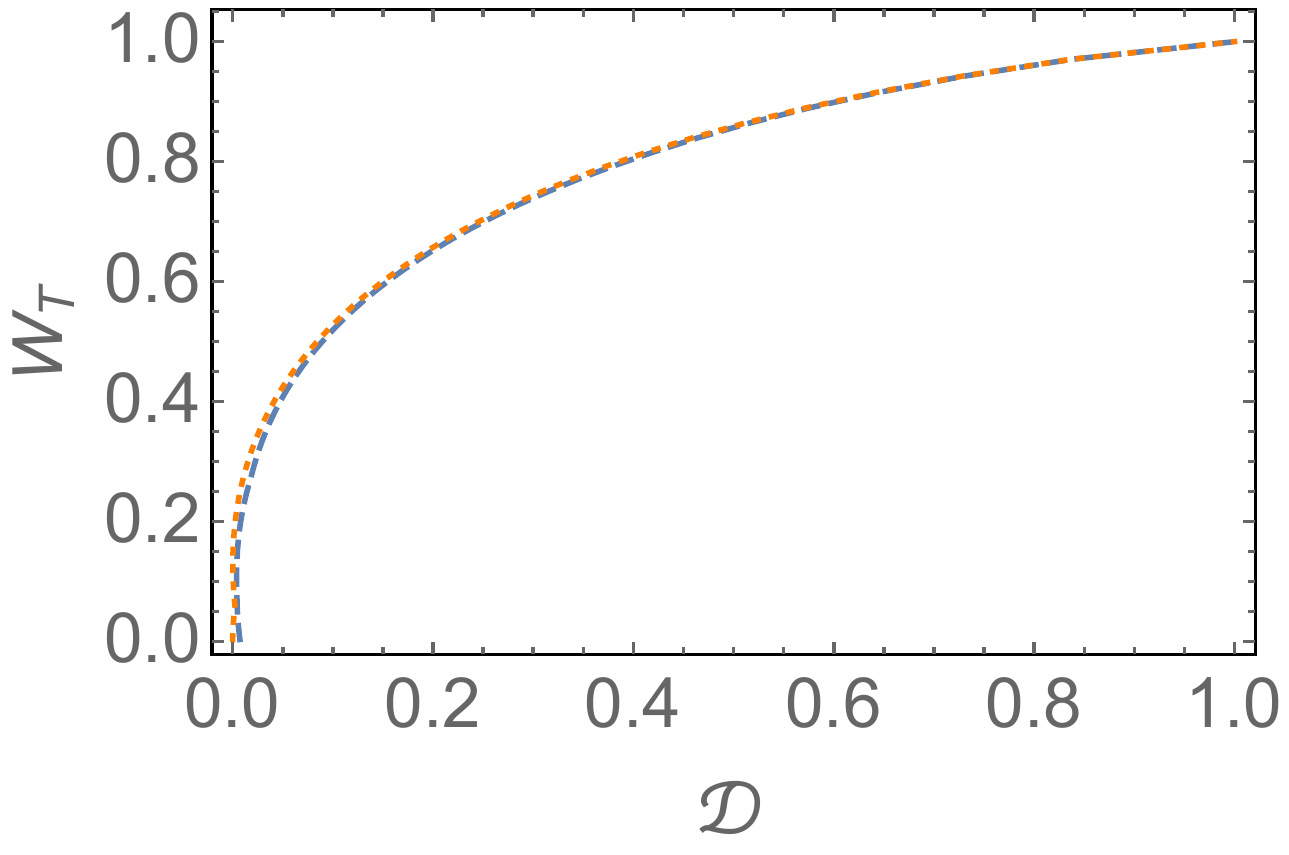}
	\caption{Local dissipation and Bell pumping. Top: Quantum discord (dashed curved) and concurrence (dotted curve) of \eqref{premeasurementstate} as a function of the parameter $\gamma$ with parameters $\bar{n}=1,J_1=1,J_2=2$.
		Bottom: Total work versus discord of the steady state of the system after connecting with the cold before measurement is performed \eqref{premeasurementstate}, (dashed curve) and after measurement \eqref{postmeasurementstate} (dotted curve), with the same parameters as in the top panel.}
	\label{WorkEtaVsDiscordMixed}
\end{figure}

 Unlike the previous case, this system is not entangled for all $\gamma>0$, resembling the situation obtained in Sec.~\ref{Local Hamiltonian}. Thus to measure the level of coherence in the system we return again to the quantum discord which is non-zero for all $\gamma>0$ (see Fig.~\ref{WorkEtaVsDiscordMixed}).

As shown in Fig.~\ref{WorkEtaVsDiscordMixed}, we can relate the work produced by the machine with the quantum discord of the steady state of the working medium after connecting to the cold bath.
 Another difference between this and the previous example is that the steady state  can no longer be written in terms of its energy eigenvectors, hence $\rho_S\neq \rho^P$. The general form of the density matrix after reaching the steady state of the Bell pump and the local dissipator has the form:
\begin{equation}\label{premeasurementstate}
\rho=
\begin{pmatrix}
r_{11}& 0 & 0 &r_{14} \\
0& r_{22} & r_{23} & 0 \\
0& r_{23}* & r_{33} & 0\\
r_{14}*& 0 & 0 & r_{44}
\end{pmatrix}\;.
\end{equation}
Analytical expressions of all  coefficients $r_{ij}$ can be found explicitly but are not reported due to their length.  The state after the measurement is
\begin{equation}\label{postmeasurementstate}
\rho_m=\frac 12
\begin{pmatrix}
r_{11}+r_{44}& 0 & 0 &2{\rm Re}[r_{14}] \\
0& r_{22}+r_{33} & 2{\rm Re}[r_{23}] & 0 \\
0& 2{\rm Re}[r_{23}] & r_{22}+r_{33} & 0\\
2{\rm Re}[r_{14}]& 0 & 0 &r_{11}+r_{44}
\end{pmatrix}.
\end{equation}
We see that the measurement only destroys the imaginary part of the coherences leaving the real part intact. This is the cause of the slight disparity in the discord between the projected and the nonprojected states  shown in Fig.~\ref{WorkEtaVsDiscordMixed}, which,however, rapidly disappears as $\gamma$ increases. As a consequence, also for this engine the observed relation between quantum correlations and energetic performances is robust against the measurement process.

\section{Summary}
\label{sec:summary}
In conclusion, we have presented several designs of quantum thermal engines whose working substance is made of two interacting qubits. We have modelled the interaction with the baths through Lindblad master equations that do not bring the system necessarily to equilibrium. This is not necessarily in contradiction with thermodynamic laws as long as one accounts for the extra resources necessary to maintain such non-equilibrium reservoirs. 

Interestingly, we have shown that in the case of common baths, it is possible to make a direct link between the work and the quantum correlations, entanglement and discord, produced during the cycle.  In this respect, our work contributes to the debate on whether or not quantum correlations are helpful  in the performance of quantum work engines. The models we have considered show that there is not, however,  a universal connection, and if it exists, it relies on the specific models we design.

Finally, given the simplicity of the model and the analytical results found in our work, we note  the possibility of realising such two-qubit engines in several experimental platforms including nuclear magnetic resonance, trapped ions, photonic systems, ultracold atoms and superconducting circuits.

\acknowledgements
GDC acknowledges the KITP program ``Thermodynamics of Quantum Systems: Measurement, Engines, and Control'' 2018, where part of this work has been done. This research was supported in part by the National Science Foundation under Grant No. NSF PHY-1748958.

\bibliography{biblio}  
\clearpage
\newpage
\appendix  
\section*{Appendix}

\section{Finite time cycles}
\label{app:partial}
The calculations in Sec.~\ref{Loc_A} assume that the system will be in contact with the heat reservoirs for a time sufficient for the system to reach steady state. In a practical implementation, the time $\tau$ in which the system interacts with the reservoir would be finite. We  now calculate the effect of partial thermalisation on the performance of the engine with equal temperatures considered in Sec.~\ref{Loc_A}.

\begin{figure}[t!]
	\centering
	\includegraphics[height=5cm]{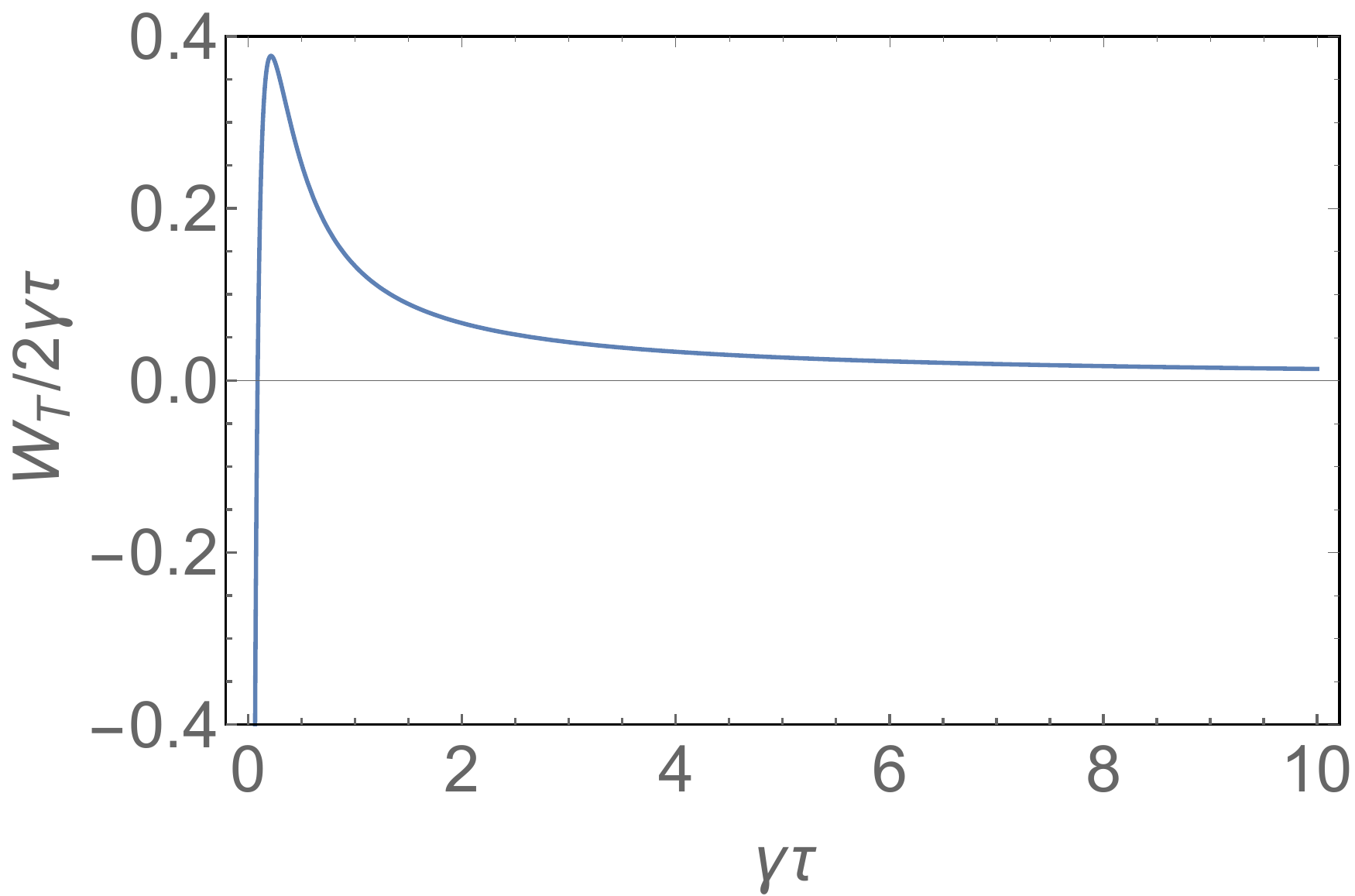}
	\caption{Plot of the power $W_{T}/2\gamma\tau$ (in units of $\gamma$) vs the duration $\tau$ of the heating and cooling strokes. Parameters: $n_{C}=1,n_{H}=2,B_1=1,B_2=2$.}
	\label{fig:power}
\end{figure}

We begin by pointing out that, in this model, the eigenvectors of the system's Hamiltonian do not change as the magnetic field is ramped. Thus the state of the system does not change during the compression and work steps (although its energy does change). We thus assume that they are done as quenches in a time negligible for our analysis. We also assume that, during the heating and cooling strokes, the system interacts, respectively, with the hot and cold baths for the same amount of time $\tau$ giving a total cycle time of $2\tau$. For the limiting cycle we find the heat and work contributions for each stroke (see Sec.~\ref{Loc_A}):
\begin{eqnarray}
W_{1}&=&\dfrac{2 (\text{B}_1-\text{B}_2)}{2 \text{n}_C+1} \left[\dfrac{2 (\text{n}_C-\text{n}_H) e^{-2 \gamma  (2 \text{n}_C+1) \tau}}{2 \text{n}_H+1}+1\right],\\
Q_{1}&=&\dfrac{4 \text{B}_2 (\text{n}_C-\text{n}_H) \Gamma}{(2 \text{n}_C+1) (2 \text{n}_H+1)},\\
W_{2}&=&\dfrac{2 (\text{B}_1-\text{B}_2)}{2 \text{n}_C+1} \left[\dfrac{2 (\text{n}_C-\text{n}_H) e^{-2 \gamma  (2 \text{n}_H+1) \tau}}{2 \text{n}_C+1}-1\right],\\
Q_{2}&=&\dfrac{4 \text{B}_1 (\text{n}_H-\text{n}_C) \Gamma}{(2 \text{n}_C+1) (2 \text{n}_H+1)}.
\end{eqnarray}
where 
\begin{equation}
\Gamma=e^{-4 \gamma   (\text{n}_C+\text{n}_H+1)\tau} \left(-e^{4 \gamma   (\text{n}_C+\text{n}_H+1)\tau}+e^{2 \gamma  (2 \text{n}_C+1) \tau}+e^{2 \gamma  (2 \text{n}_H+1) \tau}\right).
\end{equation}
We see that the amount of work and heat exchanged assuming a finite time thermalisation has a similar form to those assuming total thermalisation with the addition of time dependent exponential factors. In the limit of $\tau \rightarrow \infty$ the steady state results are returned. The total work is 
\begin{equation}
W_{T}=\dfrac{4 (\text{B}_1-\text{B}_2) (\text{n}_H-\text{n}_C) \Gamma}{(2 \text{n}_C+1) (2 \text{n}_H+1)},
\end{equation}
which is the same as Eq~\eqref{enegineperfomance} but with the additional factor of $\Gamma$. As this factor of $\Gamma$ is also present in the heat input, the efficiency of the system is still the standard Otto efficiency $\eta= 1-B_1/B_2$. However the total amount of work produced depends non trivially on $\tau$. In Fig.~\ref{fig:power} we show the total power $W_{T}/2\tau$ as a function of $\tau$. We find that for $\tau\to\infty$, the power decays exponentially to 0 as expected. There exists a value of $\tau$ below which there is no work production and the power becomes negative. Finally, there is an optimal value of $\tau$ at which the output power is maximum. No analytical expressions for these special values of $\tau$ can be obtained.

\section{Steady state coefficients for different qubit temperatures}
\label{app1}
Here we provide the analytical formula for the steady state, \eqref{unequalsteadystate}, with unequal temperatures:

\begin{widetext}
\begin{align*}
\alpha=&\left(\text{$\gamma_a$}+\text{$\gamma_b$}+2 \text{$\gamma_a$} \bar{n}_a+2 \text{$\gamma_b$} \bar{n}_b\right){}^2 \left\{\text{$\gamma_a$} \text{$\gamma_b$}+4 J^2+2 \text{$\gamma_a$} \text{$\gamma_b$} \left[\bar{n}_b+\bar{n}_a \left(2 \bar{n}_b+1\right)\right]\right\},\\
r_{11}=&4 \text{$\gamma_a$}^2 \bar{n}_a^2 \left[J^2+\text{$\gamma_b$} \bar{n}_b \left(\text{$\gamma_a$}+\text{$\gamma_b$}+2 \text{$\gamma_b$} \bar{n}_b\right)\right]+\text{$\gamma_a$} \text{$\gamma_b$} \bar{n}_b \bar{n}_a \left[(\text{$\gamma_a$}+\text{$\gamma_b$})^2+8 J^2+4 \text{$\gamma_b$} \bar{n}_b \left(\text{$\gamma_a$}+\text{$\gamma_b$}+\text{$\gamma_b$} \bar{n}_b\right)\right]
\\&+4 \text{$\gamma_b$}^2 J^2 \bar{n}_b^2+4 \text{$\gamma_a$}^3 \text{$\gamma_b$} \bar{n}_b \bar{n}_a^3,\\
r_{22}=&4 \text{$\gamma_a$}^2 \bar{n}_a^2 \left[\text{$\gamma_b$} (\text{$\gamma_a$}+\text{$\gamma_b$})+J^2+\text{$\gamma_b$} \bar{n}_b \left(\text{$\gamma_a$}+3 \text{$\gamma_b$}+2 \text{$\gamma_b$} \bar{n}_b\right)\right]
\\&+\text{$\gamma_a$} \bar{n}_a \left(\text{$\gamma_a$}+\text{$\gamma_b$}+2 \text{$\gamma_b$} \bar{n}_b\right) \left[\text{$\gamma_b$} (\text{$\gamma_a$}+\text{$\gamma_b$})+4 J^2+\text{$\gamma_b$} \bar{n}_b \left(\text{$\gamma_a$}+3 \text{$\gamma_b$}+2 \text{$\gamma_b$} \bar{n}_b\right)\right]
\\&+4 \text{$\gamma_b$} J^2 \bar{n}_b \left(\text{$\gamma_a$}+\text{$\gamma_b$}+\text{$\gamma_b$} \bar{n}_b\right)+4 \text{$\gamma_a$}^3 \text{$\gamma_b$} \left(\bar{n}_b+1\right) \bar{n}_a^3,\\
r_{33}=&4 \text{$\gamma_b$}^2 \bar{n}_b^2 \left[\text{$\gamma_a$} (\text{$\gamma_a$}+\text{$\gamma_b$})+J^2+\text{$\gamma_a$} \bar{n}_a \left(3 \text{$\gamma_a$}+\text{$\gamma_b$}+2 \text{$\gamma_a$} \bar{n}_a\right)\right]
\\&+\text{$\gamma_b$} \bar{n}_b \left(\text{$\gamma_a$}+\text{$\gamma_b$}+2 \text{$\gamma_a$} \bar{n}_a\right) \left[\text{$\gamma_a$} (\text{$\gamma_a$}+\text{$\gamma_b$})+4 J^2+\text{$\gamma_a$} \bar{n}_a \left(3 \text{$\gamma_a$}+\text{$\gamma_b$}+2 \text{$\gamma_a$} \bar{n}_a\right)\right]
\\&+4 \text{$\gamma_a$} J^2 \bar{n}_a \left(\text{$\gamma_a$}+\text{$\gamma_b$}+\text{$\gamma_a$} \bar{n}_a\right)+4 \text{$\gamma_a$} \text{$\gamma_b$}^3 \left(\bar{n}_a+1\right) \bar{n}_b^3,\\
r_{44}=&(\text{$\gamma_a$}+\text{$\gamma_b$})^2 \left(\text{$\gamma_a$} \text{$\gamma_b$}+4 J^2\right)+\text{$\gamma_a$} \bar{n}_a \left\{(\text{$\gamma_a$}+\text{$\gamma_b$}) \left(\text{$\gamma_b$} (5 \text{$\gamma_a$}+\text{$\gamma_b$})+8 J^2\right)+\text{$\gamma_b$} \bar{n}_b \left[5 \text{$\gamma_a$}^2+18 \text{$\gamma_a$} \text{$\gamma_b$}+5 \text{$\gamma_b$}^2+8 J^2+4 \text{$\gamma_b$} \bar{n}_b \left(3 \text{$\gamma_a$}+2 \text{$\gamma_b$}+\text{$\gamma_b$} \bar{n}_b\right)\right]\right\}
\\&+4 \text{$\gamma_a$}^2 \bar{n}_a^2 \left[\text{$\gamma_b$} (2 \text{$\gamma_a$}+\text{$\gamma_b$})+J^2+\text{$\gamma_b$} \bar{n}_b \left(2 \text{$\gamma_a$}+3 \text{$\gamma_b$}+2 \text{$\gamma_b$} \bar{n}_b\right)\right]
\\&+\text{$\gamma_b$} \bar{n}_b \left[(\text{$\gamma_a$}+\text{$\gamma_b$}) \left(\text{$\gamma_a$} (\text{$\gamma_a$}+5 \text{$\gamma_b$})+8 J^2\right)+4 \text{$\gamma_b$} \bar{n}_b \left(\text{$\gamma_a$}^2+2 \text{$\gamma_a$} \text{$\gamma_b$}+J^2+\text{$\gamma_a$} \text{$\gamma_b$} \bar{n}_b\right)\right]+4 \text{$\gamma_a$}^3 \text{$\gamma_b$} \left(\bar{n}_b+1\right) \bar{n}_a^3,\\
r_{23}=&2 \text{$\gamma_a$} \text{$\gamma_b$} J \left(\bar{n}_a-\bar{n}_b\right).
\end{align*}
For the case in which the second qubit is attached to a zero temperature bath, $\bar{n}_b=0$ this simplifies to
\begin{equation}
\begin{matrix}
\alpha=&[(1+2\bar{n}_a)\gamma_a+\gamma_b]^2[4J^2+(1+2\bar{n}_a)\gamma_a\gamma_b],\\[1.5ex]
r_{11}=&4J^2\bar{n}_a^2\gamma_a^2,\\[1.5ex]
r_{22}=&\bar{n}_a\gamma_a4J^2(1+\bar{n}_a)\gamma_a+[4J^2+(1+2\bar{n}_a)^2\gamma_a^2]\gamma_b+2(1+2\bar{n}_a)\gamma_a\gamma_b^2+\gamma_b^3,\\[1.5ex]	
r_{33}=&4J^2\bar{n}_a\gamma_a[(1+\bar{n}_a)\gamma_a+\gamma_b],\\[1.5ex]
r_{44}=&4J^2(1+\bar{n}_a)^2\gamma_a^2+(1+\bar{n}_a)\gamma_a[8J^2+(1+2\bar{n}_a)^2\gamma_a^2]\gamma_b+2\gamma_b^2[2J^2+(1+\bar{n}_a)(1+2\bar{n}_a)\gamma_a^2]+(1+\bar{n}_a)\gamma_a\gamma_b^3,\\[1.5ex]
r_{23}=&2J\bar{n}_a\gamma_a\gamma_b.
\end{matrix}
\end{equation}
From \eqref{unequalwork} we have: 
\begin{equation}
W_T
=(B_2-B_1)(r_{11}^H-r_{11}^C+r_{44}^C-r_{44}^H),
\end{equation}
Now, denoting  $\alpha_i= [(1+2\bar{n_i})\gamma_a+\gamma_b]^2(4J^2+(1+2\bar{n_i})\gamma_a\gamma_b)$ and using $\bar n_C $ and $\bar n_H$ for the cold and hot temperatures, respectively, we obtain
\begin{align*}
W_T= \alpha_C^{-1}\Big\{4J^2\bar n_C^2\gamma_a^2-4J^2(1+\bar n_C)^2\gamma_a^2+(1+\bar n_C)\gamma_a[8J^2+(1+2\bar n_C)^2\gamma_a^2]\gamma_b+ 
\\
 2[2J^2+(1+\bar n_C)(1+2\bar n_C)\gamma_a^2]\gamma_b^2+(1+\bar n_C)\gamma_a\gamma_b^3\Big\}-\\
\alpha_H^{-1}\left\{4J^2\bar n_H^2\gamma_a^2-4J^2(1+\bar n_H)^2\gamma_a^2+(1+\bar n_H)\gamma_a[8J^2+(1+2\bar n_H)^2\gamma_a^2]\gamma_b+
\right .
\\
\left .2[2J^2+(1+\bar n_H)(1+2\bar n_H)\gamma_a^2]\gamma_b^2+(1+\bar n_H)\gamma_a\gamma_b^3\right\}.
\end{align*}
\end{widetext}
\section{Quantum discord}
\label{app:discord}
 To measure the level of quantum correlations in the system we use the quantum discord of the steady state \cite{PhysRevA.81.042105,PhysRevLett.88.017901, henderson2001classical, 0034-4885-81-2-024001}, which is calculated as follows.
For any system the total amount of correlation present is equal to the mutual information of the system:,
\begin{equation}
\mathcal{I}(\rho^{AB})= S(\rho^A)+S(\rho^B)-S(\rho^{AB}),
\end{equation}
where $\rho^{AB} $ is the density matrix of the complete system and $\rho^{A} $ [$\rho^{B} $]  is the state of subsystem A[B]. $S(\rho) $ is the von Neumann entropy, $S(\rho)=-{\rm Tr}\rho\log_2\rho$. This mutual information is composed of the classical correlations and the quantum correlations, the so-called quantum discord. The classical correlations can be calculated by the maximum information that can be obtained by measuring one of the subsystems, \begin{equation}
\mathcal{J}(\rho^{AB})=S(\rho^{B})-\min_{\Pi_A}\sum_{i}^{N}p_i S(\rho_i),
\end{equation}
where $p_i ={\rm Tr}[\Pi_i\rho^{AB}\Pi_i]$ and $\rho_i= {\rm Tr}_A[\Pi_i\rho^{AB}\Pi_i]$. The minimisation is done over all possible set of measurements $\Pi$. The discord is then just the difference between the mutual information  and the classical correlations:
 \begin{equation}
 \mathcal{D}(\rho^{AB})=\mathcal{I}(\rho^{AB})-\mathcal{J}(\rho^{AB}).
 \end{equation}

\end{document}